\documentclass[a4paper]{iopart}
\pdfoutput=1
\usepackage[english]{babel}
\usepackage{amssymb}
\usepackage{amsthm}
\usepackage{bm}
\usepackage{color}
\usepackage{iopams}
\usepackage{latexsym}
\usepackage{times}
\usepackage{url}
\usepackage{hyperref}
\hypersetup{
  pdftitle={Nested solutions of gravitational condensate stars},
  colorlinks=true,        % false: boxed links; true: coloured links
  linkcolor=blue,         % colour of internal links
  citecolor=cyan,         % colour of links to bibliography
}

\newcommand{\cf}{cf.~}
\newcommand{\ie}{i.e.,~}
\newcommand{\eg}{e.g.,~}

\RequirePackage[fleqn]{amsmath}
\usepackage{graphicx}
\usepackage[labelfont=bf, font=small, margin={6pc,0pc}]{caption}
\usepackage{enumitem}
\begin{document}

\title{Nested solutions of gravitational condensate stars}

\author{Daniel Jampolski$^1$ and Luciano Rezzolla$^{1,2,3}$}

\address{$^1$ Institut f{\"u}r Theoretische Physik, Max-von-Laue-Str.  1,
  60438 Frankfurt, Germany}
\address{$^2$ Frankfurt Institute for Advanced Studies,
  Ruth-Moufang-Str. 1, 60438 Frankfurt, Germany}
\address{$^3$ School of Mathematics, Trinity College, 17 Westland Row, Dublin 2, Ireland}
\begin{abstract}
Black holes are normally and naturally associated to the end-point of
gravitational collapse. Yet, alternatives have been proposed and a
particularly interesting one is that of gravitational condensate stars,
or gravastars~\cite{Mazur2001}. We here revisit the gravastar model and
increase the degree of speculation by considering new solutions that are
inspired by the original model of gravastars with anisotropic pressure,
but also offer surprising new features. In particular, we show that it is
possible to nest two gravastars into each other and obtain a new solution
of the Einstein equations. Since each gravastar essentially behaves as a
distinct self-gravitating equilibrium, a large and rich space of
parameters exists for the construction of nested gravastars. In addition,
we show that these nested-gravastar solutions can be extended to an
arbitrarily large number of shells, with a prescription specified in
terms of simple recursive relations. Although these ultra-compact objects
are admittedly very exotic, some of the solutions found, provide an
interesting alternative to a black hole by having a singularity-free
origin, a full matter interior, a time-like matter surface, and a
compactness $\mathcal{C}\to (1/2)^{-}$.
\end{abstract}

%==================================================================
% PAPER START:
%==================================================================

%-------------------------------------------------------------------------
\section{Introduction}
\label{sec:introduction}
%-------------------------------------------------------------------------

Despite the compelling evidence that black holes are perfectly compatible
with gravitational-wave~\cite{Abbott2016fw} and
electromagnetic~\cite{Akiyama2019_L1_etal, EHT_SgrA_PaperI_etal}
observations, their physical existence -- and the conceptual consequences
of such an existence -- still represents a challenge for modern
physics. Because of this, a vast literature about compact astrophysical
objects with properties similar to those of black holes has been
developed over the last decades (see, \eg~\cite{Cardoso2019} for a
review). It is within this large body of works that the ingenious
solution of Mazur and Mottola~\cite{Mazur2001, Mazur2004} was proposed in
2001 as the endpoint of gravitational collapse and an alternative to the
Schwarzschild solution~\cite{Schwarzschild16a}.

The gravitational vacuum condensate star, or ``gravastar'', proposed by
Mazur and Mottola comprises an infinitesimally thin shell with the most
extreme and causal equation of state (EOS) for the (isotropic) pressure
$p$ as a function of the energy density $e$~\cite{Rezzolla_book:2013},
namely, $p=e$, and a de-Sitter interior with an EOS with $p=-e$, which is
interpreted as dark energy and is often invoked as a natural explanation
of the accelerated expansion of the universe, where it appears as a
cosmological constant. The latter EOS has a long history, starting from
1966, when Sakharov concluded that it can arise when an isolated
gravitating system reaches large baryon densities~\cite{Sakharov1966}. In
the same year, Gliner~\cite{Gliner1966} suggested that such a dark-energy
state of matter with a de-Sitter metric \cite{deSitter:1917zz} might
represent the final state of gravitational collapse. From an intuitive
point of view, while the negative pressure from the de-Sitter region
seeks to expand the gravastar, the thin shell with an ultra-stiff EOS
counterbalances the expansion leading to a static but nonvacuum solution
of the Einstein equations in spherical symmetry. Less extreme EOSs are
possible if the thin shell is replaced by a thick one and an anisotropic
pressure is added (see, \eg~\cite{Chirenti2007})\footnote{Some authors
prefer to use the term gravastar only for the static, maximally compact
and infinitesimally thin-shell configuration~\cite{Mottola2023}; for
simplicity, we will also refer to solutions with a finite-size shell as
gravastars.}. The gravastar model is not alone in the category of
nonsingular black holes. Important work in this regard has been carried
out by Dymnikova~\cite{Dymnikova1992, Dymnikova2000, Dymnikova2005},
while a remarkable connection between the gravastar and the
constant-density interior Schwarzschild solution has been identified by
Mazur and Mottola~\cite{Mazur_2015}.

There are several advantages behind a gravastar solution. First, it
provides the endpoint of gravitational collapse without the central
spacelike singularity encountered in the Schwarzschild solution. Second,
it removes altogether the problem of an information paradox, since the
gravastar surface can be placed infinitesimally outside the putative
event horizon and thus is not a null but a time-like surface (compact
objects of this type are often referred to as ``nonsingular and
horizon-less'' black holes). Indeed, in the case of gravastars, the
Buchdahl bound~\cite{Buchdahl:59} for the compactness of spherically
symmetric objects, can be easily avoided through the introduction of
anisotropic transverse stresses, resulting in a gravastar within the
framework of the interior Schwarzschild solution~\cite{Mazur_2015}. While
an isolated gravastar would be very hard to distinguish from a black hole
when employing electromagnetic radiation (see, \eg~\cite{Carballo2018}),
the response to gravitational perturbations is different, thus making
gravastars distinguishable from black holes when measuring their ringdown
signal~\cite{Chirenti2007, Chirenti2016}.

At the same time, more then 20 years since their introduction -- and
  with hundreds of papers dedicated to the investigation of their
  properties (see, \eg \cite{Ray2020, Mottola2023} for some recent and
  extensive reviews) -- the genesis of gravastars is still highly
speculative and fundamentally unclear even in the most idealised
conditions. It is claimed that quantum fluctuations amplified during the
final stages of the collapse induce a vacuum phase transition, yielding
the emergence of dark energy and enabling the formation of the
shell~\cite{Chapline_2003}. However, realistic calculations demonstrating
that this process is dynamically possible are still lacking; in addition,
the astrophysical evidence of their existence is also shrouded by
doubts~\cite{Broderick2007, Chirenti2016, EHT_SgrA_PaperVI_etal}.
However, because the considerations that we will present here are even
more speculative than those that normally accompany gravastars, we will
indulge on ignoring the complications associated with the creation of
gravastars and simply assume that these solutions are not only
mathematically possible but also realisable in nature.

We here continue the exploration of static and spherically symmetric
gravitational condensate stars by increasing the degree of speculation
and constructing a new solution consisting of two gravastars nested into
each other. Again, from an intuitive point of view, it is natural to
expect that since each of the two gravastars is already in a near
hydrostatic equilibrium, the nesting of the two solutions can lead to yet
another static equilibrium. In practice, we show that a suitable but
generic prescription for the EOS of the matter in the two shells provides
a solution of the Einstein equations for two nested gravastars, namely a
``nestar''. Intrigued by this behaviour, we explore the possibility of
nesting an additional gravastar and show that this is also possible,
leading in fact to a precise recipe for introducing an arbitrary number
of nested gravastars. Hence, in the limit of infinite nested shells it is
possible to construct a nonvacuum, nonsingular, static and matter-filled
interior of a compact object whose compactness can reach values
arbitrarily close to the Schwarzschild limit.

The structure of the paper is as follows. We begin in Sec.~\ref{sec:MM}
with a concise overview of the Mazur-Mottola thin-shell gravastar model
and discuss in Sec.~\ref{sec:aniso} how it can be extended by removing
the assumption of an infinitesimal shell and by introducing a finite-size
shell with a continuous and anisotropic pressure
profile. Section~\ref{sec:nestars} is instead dedicated to the analysis
of nested solutions of gravastars for which we can ensure continuity
conditions for all relevant spacetime and fluid functions and their first
derivatives. We also explore the rich space of parameters and solutions
that can be obtained when varying the thickness and compactness of the
matter shells. We derive expressions that can be cast in terms of generic
coefficients that follow recurrent relations that we collect in
Appendix~\ref{sec:appendix}, where we also discuss the rich radial
structure of a ``nestar''. Finally, Sec.~\ref{sec:conclusion} summarises
our results and suggests possible developments.

%-------------------------------------------------------------------------
\section{Gravastars with isotropic pressure (thin shell)}
\label{sec:MM}
%-------------------------------------------------------------------------

Before entering the details of the new nested-gravastars model, it is
useful to briefly recall the thin-shell gravastar model proposed by Mazur
and Mottola~\cite{Mazur2001, Mazur2004}. Let us start from a generic,
static and spherically symmetric line element in the form
\begin{equation}
  \label{eq:MM_line_element}
  ds^{2}= -f(r)dt^{2} + \frac{dr^{2}}{h(r)} + r^{2}(d\theta^{2} +
    \sin^2(\theta) d\phi^{2})\,,
\end{equation}
and split the spacetime into three different regions marked by the radial
coordinate $r$, namely,
\begin{eqnarray}
\text{~~I.~$\,\,$de-Sitter~interior} \hspace{1.0cm} &r \in\left[0, r_{1}\right)\,,  \hspace{1.0cm}&p=-e\,,\label{eq:MM:eos1}\\
\text{~II.~~matter~shell}   &r \in\left[r_{1}, r_{2}\right]\,, &p=e\,,\label{eq:MM:eos2}\\
\text{III.~$\,$vacuum~exterior} &r \in\left(r_{2}, \infty\right)\,, &p=e=0\,.\label{eq:MM:eos3}
\end{eqnarray}

\begin{figure*}
  \centering
  \includegraphics[width=0.45\textwidth]{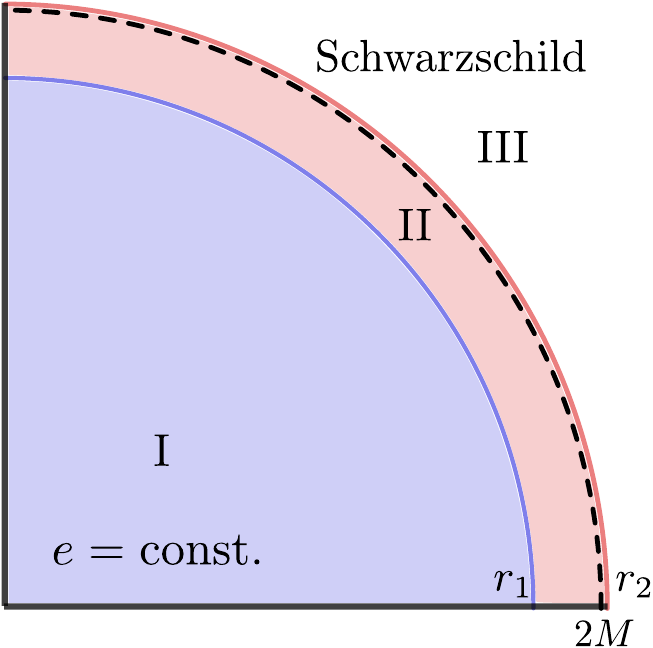}
  \caption{Illustrative representation of a gravastar. Shown with
    different colours are the three different regions of the spacetime: a
    de-Sitter region (I., blue-shaded area), the ultra-stiff
    shell (II., red-shaded area), and the Schwarzschild vacuum exterior
    (III., white area). Shown with a black dashed line is the position of
    the surface $r=2M$, with $M$ the mass of the gravastar.}
  \label{fig:gravastar_cartoon}
\end{figure*}

The Einstein equations and the momentum-conservation equation for a
perfect fluid at rest take the following form
\begin{alignat}{4}
    -G^{\,t}_{\,\,t} &= &&\frac{1}{r^2}\frac{d}{dr}[r(1-h)] &&= -8\pi
    T^{\,t}_{\,\;t} &&= 8\pi e\,, \label{eq:MM_einst1}\\[1em]
    \phantom{-}G^{\,r}_{\,\,r} &= \frac{h}{rf}&&\frac{df}{dr} +
    \frac{1}{r^2}(h-1) &&= \phantom{-}8\pi T^{\,r}_{\,\;r} &&= 8\pi p\,,
    \label{eq:MM_einst2}
\end{alignat}
\vspace{-1.0em}
\begin{equation}
    \nabla_i T^{\,i}_{\,\;r} = \frac{dp}{dr} + \frac{e+p}{2f}\frac{df}{dr} = 0\,,
    \label{eq:MM_conserv}
\end{equation}
where $G_{\mu \nu}$ and $T_{\mu \nu}$ are the Einstein and
energy-momentum tensors, respectively. The solution of
Eqs.~\eqref{eq:MM_einst1}--\eqref{eq:MM_conserv} leads, in the thin-shell
approximation $|1 - r_2/r_1| \ll 1$ and $0<h \ll 1$, to the following
metric functions
\begin{eqnarray}
  \label{eq:MM_metric_functions}
  \text{I.~de-Sitter~interior: }&
  \begin{cases}
    \begin{aligned}
      f(r) &= C(1-H_0^2r^2)\,,\\ \vspace{0.5cm}
      h(r) &= 1-H_0^2r^2\,,
    \end{aligned}
  \end{cases}\\ \nonumber\\ 
  \text{II.~matter~shell:}&
  \begin{cases}
    \begin{aligned}\label{eq:MM_metric_functions2}
      f(r) &= {w_1}f(r_1)/{w(r)}\,,\\
      h(r) &= \epsilon\,[1+w(r)]^2/w(r)\,,
    \end{aligned}
  \end{cases}\\\nonumber \\ 
  \text{III.~exterior: }&
  \begin{cases}
    \begin{aligned}\label{eq:MM_metric_functions3}
      f(r) &= 1 - {2M}/{r}\,,\\
      h(r) &= 1 - {2M}/{r}\,,
    \end{aligned}
  \end{cases}
\end{eqnarray}
where $w (r) := 8\pi r^2 \, p(r)$ is a dimensionless quantity, defined
only within the shell, with constants given by $w_1:=w(r_1)$ and
$w_2:=w(r_2)$. Enforcing continuity of the metric functions at each layer
fixes the free constants in the model, namely
\begin{align}
\epsilon &= -\ln \left(\dfrac{r_2}{r_1}\right)\left[\,\ln
  \left(\dfrac{w_2}{w_1}\right) - \dfrac{1}{w_2} +
  \dfrac{1}{w_1}\right]^{-1}\,,\\
C &=\biggl(\dfrac{1+w_2}{1+w_1}\biggr)^{2}\,, \\
H_0^2 &= \dfrac{1}{r_1^2}\left[1-\epsilon\,\dfrac{(1+w_1)^2}{w_1}\right]\,,\\
M &=\dfrac{r_2}{2}\left[1-\epsilon\,\dfrac{(1+w_2)^2}{w_2}\right]\,.
\end{align}
Finally, we make the interesting remark that if we introduce the
coordinate transformation
\vskip -1.2em%
\begin{equation}
  \label{eq:rstar}
  r^{*}=e^{\displaystyle\,\tfrac{1}{w_1}}
  \left(\frac{r}{r_1}\right)^{\displaystyle\tfrac{1}{\epsilon}}
  \frac{1}{w_1}\,,
\end{equation}
then the metric function $w(r)$ is very well approximated analytically
by the Lambert $W$ function
\begin{equation}
  w(r) = {W_0(r^{*})}^{-1}\,.
\end{equation}
When comparing to numerically computed solutions and allowing for a
finite shell thickness, denoting it by $\delta:=r_2-r_1$ and the mass of
the gravastar by $M$, this representation is very accurate in the case of
$\delta/M \ll 1$ and maintains a relative error with respect to the
numerical solution that is of the order of a couple percent in the case
of $\delta/M \simeq 1/2$. Before moving on, it is useful to remark
  that, strictly speaking, a gravastar with isotropic pressure --
  independently of whether with a thin or thick shell -- does break
  isotropy, but this happens only at the infinitesimal layer
  transitioning from the negative to the positive energy density at
  $r=r_1$ and at the infinitesimal layer transitioning from the positive
  energy density to the vacuum at $r=r_2$. These discontinuous jumps lead
  to transverse stresses, but with a delta distribution~\cite{Mazur2001,
    Mazur_2015}; everywhere else, the pressure is isotropic and this is
  why they are still considered isotropic-pressure gravastars.

\begin{figure*}
  \centering
  \includegraphics[width=0.65\textwidth]{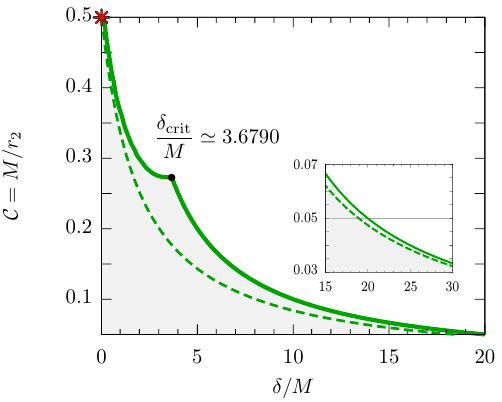}
  \caption{Space of numerical solutions for an isotropic-pressure
    gravastar shown as a function of the gravastar compactness
    $\mathcal{C}:=M/r_2$ and of the shell thickness
    $\delta:=r_2-r_1$. The solutions are found by fixing $r_2$ and then
    varying the inner radius $r_1$ and the de-Sitter energy density,
    which yield either high-compactness, thin-shell gravastars, or
    low-compactness, thick-shell gravastars. The solid green line marks
    the edge of the space of solutions beyond which an event horizon
    appears or the radius $r_1<0$, with the critical shell thickness
    $\delta_{\rm crit}$ denoting the boundary between the two different
    branches of the limit. A smaller space of solutions was found in
    Ref.~\cite{Chirenti2007} and its limit is marked with a green dashed
    line; the two limits coincide when $\delta\to0$ and
    $\delta\to\infty$, as shown in the small inset. Shown with a red
    asterisk is the isotropic-pressure (thin-shell) gravastar solution.}
\label{fig:MM:gravastar}
\end{figure*}

%=========================================================================
\subsection{Space of solutions of isotropic-pressure gravastars}
%=========================================================================

In order to explore the space of possible solutions of isotropic-pressure
gravastars, we will not only use the analytic thin-shell solution, but
also include solutions having a finite-thickness shell, which are
computed numerically, using the same EOSs in
Eqs.~\eqref{eq:MM:eos1}--\eqref{eq:MM:eos3}. Imposing that an event
horizon is not produced is equivalent to imposing that the metric
functions must remain positive. We therefore start by fixing the position
of the outer radius $r_2$, then set a maximum value for the de-Sitter
energy density and start an iteration process over all possible values of
$r_1$ and de-Sitter energy densities, checking for sign changes in the
metric functions and saving as admissible those configurations for which
this change in sign does not occur.

Figure~\ref{fig:MM:gravastar} reports with a grey-shaded area the space
of allowed solutions when expressed in terms of the shell thickness
$\delta$, and of the gravastar compactness, \ie $\mathcal{C}:=M/r_2$.
The limit of the space of solutions is marked with a solid green line and
marks the location where $g_{tt}=0$, hence the appearance of an event
horizon, or where $r_1=0$. Note that the edge of the region of allowed
solutions is characterised by a critical shell thickness
$\delta_{\rm crit}$, which distinguishes two qualitatively different
behaviours and hence branches ($\delta_{\rm crit}$ also marks a
critical compactness $\mathcal{C}_{\rm crit} \simeq 0.2718$). 
More specifically, solutions approaching the edge along lines with
$\mathcal{C} {\delta} / {M} = {\rm  const.} < 1$ and for which
$\delta > \delta_{\rm crit}$, are characterised by the hyperbolic
condition
\begin{equation}
  \label{eq:hyperb_cond}
  \mathcal{C}\frac{\delta}{M} = \frac{r_2-r_1}{r_2} \to 1 \qquad
  \Rightarrow r_1 \to 0\,,
\end{equation}
and represent what we refer to as the ``unrestricted branch'' of
solutions because, once the value for $r_2 > \delta_{\rm crit} > 2M$ is
fixed, the inner radius of the shell $r_1$ can be taken to be arbitrarily
close to zero. On the other hand, solutions approaching the edge but with
$\delta < \delta_{\rm crit}$ are such that once the value for
$\delta_{\rm crit} > r_2 > 2M$ is fixed, a minimum value for $r_1$
appears and is dependent on $r_2$; because of this, we refer to this as
the ``restricted branch''. We note that the space of solutions reported
in Fig.~\ref{fig:MM:gravastar} differs from the equivalent one shown in
Fig.~1 of Ref.~\cite{Chirenti2007}, whose limit of the space of solutions
is shown as a green dashed line.  While edges of the two spaces of
parameters coincide in the limits of $\delta \to 0$ and $\delta\to\infty$
(see inset in Fig.~\ref{fig:MM:gravastar}) the space of solutions
reported here is larger than in Ref.~\cite{Chirenti2007}, most likely
because the initialisation of the discretisation in the interior radius
and of the de-Sitter energy density is finer than that employed in
Ref.~\cite{Chirenti2007}. Finally, we mark with a red asterisk in
Fig.~\ref{fig:MM:gravastar} the isotropic-pressure (thin-shell) gravastar
solution proposed by Mazur and Mottola~\cite{Mazur2001} as a way to
remark that such solution represents a very special (and the most
extreme) solution in the space of gravastar solutions.

%-------------------------------------------------------------------------
\section{Gravastars with anisotropic pressures (thick shell)}
\label{sec:aniso}
%-------------------------------------------------------------------------

It was realised already early on by Cattoen and Visser that when moving
away from the infinitesimally thin-shell gravastar model and introducing
a more realistic finite-size shell, the pressure cannot be
isotropic~\cite{Cattoen2005}. As a result, for such anisotropic-pressure
gravastars (or thick-shell gravastars) the shell does not
have a simple ultra-stiff EOS $p=e$, but a more involved function
$p=p(e)$, which is otherwise arbitrary. Adopting the same notation as
in~\cite{Chirenti2007}, we rewrite the static and spherically symmetric
line element~\eqref{eq:MM_line_element} as
\begin{equation}
    ds^{2}= -e^{\nu(r)}dt^{2} + e^{\lambda(r)}dr^{2} + r^{2}(d\theta^{2}
    + \sin^2(\theta) d\phi^{2})\,,
    \label{eq:aniso:line_element}
\end{equation}
and consider a perfect fluid stress-energy tensor with a radial pressure
$p_r$ and a tangential pressure $p_t$ given by
$T^{\mu}_{\phantom{\mu}\nu}= \textnormal{diag}
\left(-e,p_r,p_t,p_t\right)$. The corresponding nonzero Einstein
equations then become
\begin{eqnarray}
  \label{eq:aniso:einstein1}
  e^{-\lambda(r)} &= g_{rr}(r)^{-1} = 1 - \frac{2m(r)}{r}\,,\\[0.5em]
  \label{eq:aniso:einstein2}
  \nu^{\,\prime}(r) &= \frac{2       m(r)+8\pi r^3 p_r}{r[r-2m(r)]}\,,\\[0.5em]
  \label{eq:aniso:einstein3}
  p_r^{\,\prime} &= -(e+p_r)\,\frac{m(r)+4\pi r^3 p_r}{r[r-2m(r)]} +
  \frac{2(p_t-p_r)}{r}\,,
\end{eqnarray}
where, as customary, we have defined the mass function as
\begin{equation}
  \label{eq:aniso:mass_fun_def}
  m(r) := \int_0^r 4\pi  r^2 e(r)\,dr\,,
\end{equation}
and used a prime $'$ to indicate the total radial derivative.

%=========================================================================
\subsection{Conditions on the fluid variables}
%=========================================================================

Clearly, at the interfaces between the different regions of the spacetime
we wish to impose the continuity of the energy density, its first and
second derivative, thus ensuring the corresponding continuity of both the
radial pressure (up to the second derivative) and of the tangential
pressure (up to the first derivative). This marks a difference with
respect to what proposed in ~\cite{Chirenti2007}, where the
continuity of the energy density was ensured only up to the first
derivative, thus leading to a discontinuity in the first derivative of
the tangential pressure. We therefore enforce  
\begin{eqnarray}
  \label{eq:aniso:en_den_conds_1}
  &e(0) = e(r_1) = e_1\,, \hspace{2.0cm}& e(r_2) = 0\,, \\
  \label{eq:aniso:en_den_conds_2}
  &e'(r_1) = 0 = e'(r_2)\,, & e''(r_1)=0=e''(r_2)\,.
\end{eqnarray}
We can easily satisfy these conditions via the following energy-density
function
\begin{equation}
\label{eq:e_of_r}  
  e(r) = \left\{\begin{aligned}
  \vspace{0.1em}
  &\, e_1 = {\rm const.} &\quad {\rm for} \quad & 0\leq r < r_1\,,
  \quad&&\mathrm{:~I.~interior}\,,\\
  \vspace{0.1em}
  &\, \sum^{5}_{i=0} a_i\,r^i &\quad {\rm for} \quad & r_1 \leq r \leq r_2\,,
  \quad  &&\mathrm{:~II.~shell}\,,\\
  \vspace{0.1em}
  &\quad 0&\quad {\rm for} \quad & r_2 < r\,,
  \quad&&\mathrm{:~III.~exterior}\,,
  \end{aligned}
  \right.
\end{equation}
where the six coefficients $a_0-a_5$ can be computed after a bit of
algebra and are given by
\begin{alignat}{3}
  a_5&=-\frac{6e_1}{(r_2-r_1)^5}\,, \quad
  &a_4&=\frac{15e_1(r_1+r_2)}{(r_2-r_1)^5}\,,\nonumber\\[0.5em]
  a_3&= -\frac{10e_1(r_1^2+4r_1r_2+r_2^2)}{(r_2-r_1)^5}\,, \quad
  &a_2&=\frac{30e_1 r_1 r_2(r_1+r_2)}{(r_2-r_1)^5}\,,\\[0.5em]
  a_1&=- \frac{30e_1 r_1^2r_2^2}{(r_2-r_1)^5}\,, \quad
  &a_0&=\frac{e_1 r_2^3(10r_1^2-5r_1r_2+r_2^2)}{(r_2-r_1)^5}\,.\nonumber
\end{alignat}
To obtain the de-Sitter interior with constant energy density $e_1$ in
terms of the total mass $M := m(r_2)$ we write
\begin{equation}
    e_1 = \frac{21M}{\pi}\left(5\,r_1^3 + 9\,r_1^2\,r_2 + 9\,r_1\,r_2^2
    + 5\,r_2^3\right)^{-1}\,.
\end{equation}
Next, while the EOS for the radial pressure $p_r$ in the interior and exterior
follows the prescription of \eqref{eq:MM:eos1} and \eqref{eq:MM:eos3},
we employ the EOS for the radial pressure in the shell as suggested in
Ref.~\cite{Mbonye2005} and which is a nonlinear function of the
energy density with both a quadratic and a quartic term, namely,
\begin{equation}
  \label{eq:aniso:eos}
  p_r(e)= \left(\frac{e^2}{e_1}\right)
  \Biggl[\alpha-(1+\alpha)\left(\frac{e}{e_1}\right)^2\Biggr]\,,
\end{equation}
where the constant $\alpha$ can be constrained by requiring that the
sound speed $c_s^2 := dp_r/de$ is always subluminal, which yields
\begin{equation}
  \alpha \leq \alpha_{\rm crit} := \frac{3}{4}
  \Biggl[\sqrt[3]{4-2\sqrt{2}}+\sqrt[3]{2(2+\sqrt{2})}\,\Biggr] \simeq
  2.2135\,.
\end{equation}
Admittedly, the EOS \eqref{eq:aniso:eos} is not particularly realistic;
however, it satisfies the basic properties of a gravastar, that is,
$p_r(r_1) = -e$ and $p_r(r_2) = 0$, and provides a convenient closure
relation for the system of equations when considering $\alpha =
\alpha_{\rm crit}$. It further fulfills the weak energy condition (WEC)
and the dominant energy condition (DEC), and the corresponding sound
speed does not have a maximum at $e=0$, as discussed later on in
\ref{conditions:nestar_eos}, these conditions come from
Ref.~\cite{Mbonye2005}. To complete the system, we still need an
expression for the tangential pressure, which can be computed after
rearranging Eq.~\eqref{eq:aniso:einstein3} as
\begin{equation}
    p_t = p_r + \frac{r}{2}\,p_r^{\prime}
    +\frac{1}{2}(e+p_r)\,\Biggl[\frac{m(r)+4\pi r^3
        p_r}{r-2m(r)}\Biggr]\,.
    \label{eq:anisotropic_tangential_pressure}
\end{equation}

%=========================================================================
\subsection{Conditions on the metric functions}
%=========================================================================

Of course, suitable continuity conditions need to be met and enforced
also for the metric functions.  It is not difficult to deduce that $m(r)$
and $m^{\prime}(r)$ are continuous because of our choice of the energy
density $e(r)$, so that also the metric function $g_{rr}$, as
defined in Eq.~\eqref{eq:aniso:einstein1}, is continuous. On the
other hand, we must allow for an integration constant for the metric
function $g_{tt} = -e^{\nu(r)}$ with
\begin{equation}
    \nu(r)=\int^r_0 \dfrac{2m(r)+8\pi r^3
      p_r}{r[r-2m(r)]}\,dr +\nu_0\,.
\end{equation}
To determine the integration constant, we impose continuity at the
surface of the gravastar with $g_{tt}(r_2) = -(1-2M/r_2)$, which yields
\begin{equation}
    g_{tt} = -\left(1-\frac{2M}{r_2}\right) e^{\Gamma(r)-\Gamma(r_2)}\,,
\end{equation}
where
\begin{equation}
    \Gamma(r):=\int^r_0 \dfrac{2m(r)+8\pi r^3 p_r }{r[r-2m(r)]}\,dr\,.
\end{equation}
Reference~\cite{Chirenti2007} carried out a systematic analysis to 
show that the free parameters $M$, $r_1$ and $r_2$ cannot
be chosen arbitrarily, but in such a way that $g_{rr}$ must always be
positive to prevent the appearance of event horizons, thus requiring that
\begin{equation}
  \label{eq:aniso:no_horizon_inequality} 
  m(r) < \frac{r}{2} \qquad \forall\,r\,.
\end{equation}

\begin{figure*}
  \centering
  \includegraphics{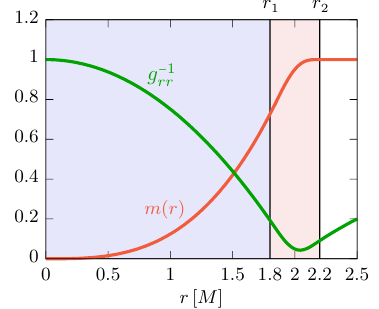}
  \includegraphics{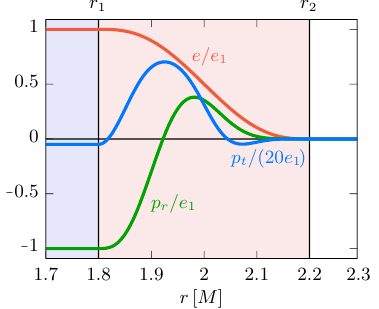}
  \caption{Metric functions (left panel) and fluid pressures and energy
    density (right panel) for a representative anisotropic-pressure
    gravastar with a mass $M$. In analogy with
    Fig.~\ref{fig:gravastar_cartoon}, the blue (red) shading refers to
    the de-Sitter interior (matter shell); the inner radius is
    $r_1=1.8\,M$ and the outer radius is $r_2=2.2\,M$. The tangential
    pressure is scaled by a factor of $1/20$ for better visualisation,
    but provides the dominant contribution in the Einstein equations.}
  \label{fig:anisotropic:en_pres_metr_mass}
\end{figure*}

Figure~\ref{fig:anisotropic:en_pres_metr_mass} provides a useful example
of the radial behaviour of the metric quantities (left panel) and of the
fluid quantities (right panel) for a representative anisotropic-pressure
gravastar with a mass $M$. The inner radius is $r_1=1.8\,M$ and the outer
radius is $r_2=2.2\,M$; note that the tangential pressure is scaled by a
factor of $1/20$ for better visualisation, but effectively provides the
dominant contribution in the Einstein equations.

%=========================================================================
\subsection{Space of solutions of anisotropic-pressure gravastars}
%=========================================================================

\begin{figure*}
  \centering
  \includegraphics[width=0.65\textwidth]{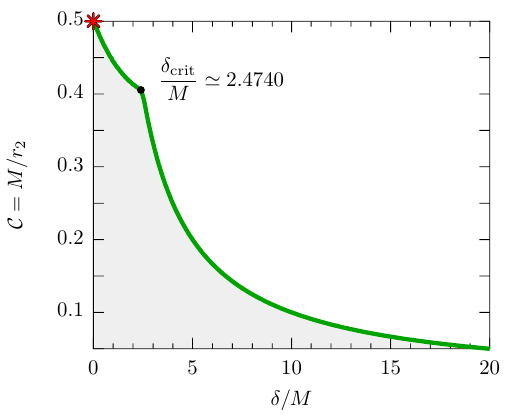}
  \caption{The same as in Fig.~\ref{fig:MM:gravastar} but for an
    anisotropic-pressure gravastar. Shown with a red asterisk is the
    isotropic-pressure (thin-shell) gravastar solution.}
\label{fig:anis:sol_space}
\end{figure*}

Also when constructing gravastars with an anisotropic pressure, the space
of possible solutions is limited in terms of the compactness
$\mathcal{C}:=M/r_2$ and shell thickness $\delta:=r_2-r_1$.
Fig.~\ref{fig:anis:sol_space} shows this behaviour, and the corresponding
data has been obtained by following the same approach employed for the
isotropic thin-shell gravastar, where the solid green line marks the edge
of the allowed solutions, denoting with $\delta_{\rm crit}\simeq
2.4740\,M$ the critical shell thickness with a corresponding critical
compactness of $\mathcal{C}_{\rm crit}\simeq0.4042$, dividing the edge
again into a restricted branch with $\delta<\delta_{\rm crit}$ and an
unrestricted branch with $\delta>\delta_{\rm crit}$
\footnote{We note that the critical thickness reported in
Fig.~\ref{fig:anis:sol_space} differs from the equivalent one shown in
Fig.~4 of Ref.~\cite{Chirenti2007}, whose value is slightly lower since a
cubic energy-density polynomial was instead employed there.}. The
existence of the restricted branch is essentially the result of our
choice for the energy-density function~\eqref{eq:e_of_r} and of the
ability to prevent the formation of an event horizon by satisfying the
condition~\eqref{eq:aniso:no_horizon_inequality} [we recall that
  different choices of $e(r)$ automatically map into different choices of
  the function $m(r$) via the
  definition~\eqref{eq:aniso:mass_fun_def}]. Indeed, apart from the
two-branches case considered here, other and suitable choices of the
function $e(r)$ would lead to either a single unrestricted branch
following the condition~\eqref{eq:hyperb_cond} or to a single restricted
branch. It is possible to examine this behaviour by looking at the limit
of $r_1 \to 0$, inspecting the energy density, mass function and
condition~\eqref{eq:aniso:no_horizon_inequality}, noting whether the
condition is fulfilled for all compactnesses or not. Solutions cannot go
beyond the hyperbole~\eqref{eq:hyperb_cond}, as that would imply negative
$r_1$; also in this case, a red asterisk marks the isotropic-pressure
(thin-shell) gravastar. A more detailed discussion of the radial
structure of the gravastar solutions along the edge of equilibria can be
found in Appendix~\ref{appendix:radstruc}.

%

%-------------------------------------------------------------------------
\section{Nestars}
\label{sec:nestars}
%-------------------------------------------------------------------------

Having recalled the construction of anisotropic-pressure gravastars, we
can now proceed with the same basic formalism to develop instead a
multi-layered compact object composed of \textit{two} nested gravastars,
that is, a ``nestar''. In practice, we now split the spherically
symmetric and static spacetime into the following five regions:
\begin{align}
& \text{I.~~~$\,\,$de-Sitter~shell}& &r
  \in\left[0, r_{1}\right)\,,& \hspace{0.0cm}& p=-e_1\,,\label{eq:N:eos1}\\
&\text{II.~~~matter~shell} & &r
    \in\left[r_{1}, r_{2}\right]\,,& \hspace{0.0cm}&
    p=p(e)\,,\label{eq:N:eos2}\\
& \text{III.~$\,$de-Sitter-shell}& &r
    \in\left(r_{2}, r_3\right)\,,& \hspace{0.0cm}&
      p=-e_3\,,\label{eq:N:eos3}\\
& \text{IV.~~matter~shell} & &r
  \in\left[r_{3}, r_{4}\right]\,,& \hspace{0.0cm}&
  p=p(e)\,,\label{eq:N:eos4}\\
& \text{V.~~~vacuum~exterior}& &r
  \in\left(r_{4}, \infty\right)\,, &\hspace{0.0cm}
  &p=e=0\,.\label{eq:N:eos5}\hspace{1.0cm}
\end{align}

\begin{figure*}
  \centering
  \includegraphics[width=0.45\textwidth]{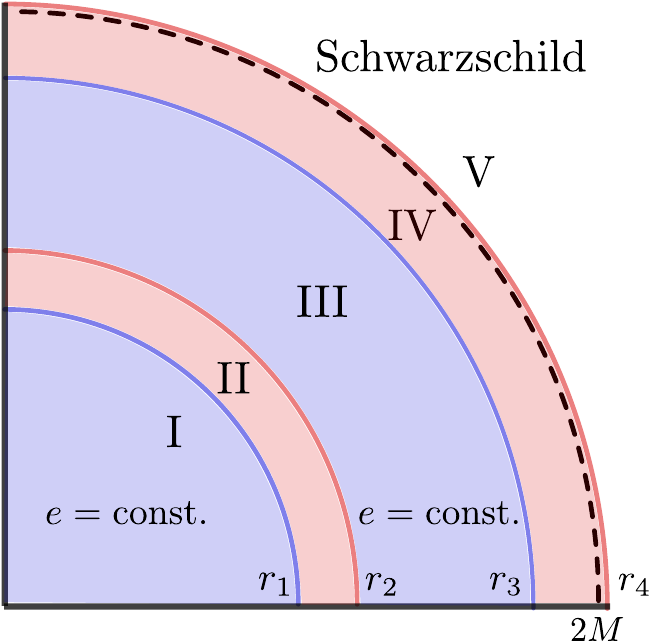}
  \caption{Illustrative representation of a nestar. Note the addition of
    regions III and IV, which represent a replica of regions I and II,
    before the spacetime is smoothly joined with the Schwarzschild
    portion V.}
\label{fig:nestar_cartoon}
\end{figure*}

Obviously, this construction would not work if one had in mind the
nesting of two stratified ordinary stars, as the addition of the second
star would completely change any hydrostatic equilibrium (the new
shells would just increase the gravitational field as gravity is
attractive for both). However, as discussed in the introduction, it is
reasonable to expect that, in the case of gravastars, this construction
could lead to a new equilibrium as each of the two gravastar can be
considered in near hydrostatic equilibrium, so that nesting them into
each other should be possible. Figure \ref{fig:nestar_cartoon} provides
an illustrative representation of a nestar, where the addition of regions
III and IV should be noted, which represent a replica of regions I
and II, before the spacetime is smoothly joined with the Schwarzschild
portion V. Finally, as we will see below, this intuition of nesting
gravastars as in a ``matryoshka doll'' -- a collection of wooden dolls
diminishing in size and nested within each other -- is correct in the
case of anisotropic-pressure (thick-shell) gravastars and incorrect in
the case of isotropic-pressure (thin-shell) gravastars.

%=========================================================================
\subsection{No-go theorem for an isotropic-pressure (thin-shell) nestar}
\label{sec:MM_impossibility}
%=========================================================================

We start by proceeding in analogy with what done in the case of the
construction of the gravastar layered spacetime and split the nestar
spacetime into the following layers [\cf
  Eqs.~\eqref{eq:MM_metric_functions} -- \eqref{eq:MM_metric_functions3}]
\begin{eqnarray}
  \label{eq:nestar_structure_1}
  \text{I.~~~1st~de-Sitter~shell:~~ }&
  \begin{cases}
    \begin{aligned}
      f(r) &= C(1-H_0^2r^2)\,,\\[1.0em]
      h(r) &= 1-H_0^2r^2\,,
    \end{aligned}
  \end{cases}
  \\[1.0em]
  \text{II.~~1st~matter-shell: ~~~}&
  \begin{cases}
    \begin{aligned}
      f(r) &= {w_1}\,f(r_1)/{w(r)}\,,\\[1em]
      h(r) &= \epsilon\,{[1+w(r)]^2}/{w(r)}\,,
    \end{aligned}
  \end{cases}
\end{eqnarray}

\begin{eqnarray}
  \label{eq:nestar_structure_2}
  \text{III.~2nd~de-Sitter~shell: }&
  \begin{cases}
    \begin{aligned}
      f(r) &= D(1-\tilde{H}_0^2 r^2)\,,\\[1.0em]
      h(r) &= 1-\tilde{H}_0^2 r^2\,,
    \end{aligned}
  \end{cases}
  \\[1.0em]
  \text{IV.~2nd~matter~shell: }&
  \begin{cases}
    \begin{aligned}
      f(r) &= {w_3}\,f(r_3)/{w(r)}\,,\\[1em]
      h(r) &= \gamma\,{[1+w(r)]^2}/{w(r)}\,,
    \end{aligned}
  \end{cases}
  \\[1.0em]            
  \label{eq:nestar_structure_3}
  \text{V.~$\,\,$exterior: }&
  \begin{cases}
    \begin{aligned}
      f(r) &= 1 - {2M}/{r}\,,\\[1em]
      h(r) &= 1 - {2M}/{r}\,,
    \end{aligned}
  \end{cases}
\end{eqnarray}
where $\epsilon, \gamma, C, D, H_0, \tilde{H}_0$, and $M$ are constants
to be determined via continuity conditions on the metric functions, while
$r_i$ and $w_i:=w(r_i)$, with $i=1,...,4$, are free parameters.

As dictated by the thin-shell approximation, the metric functions $h(r)$
in the matter-shell regions II and IV have to go to zero in this limit,
namely,
\begin{equation}
    h_{\rm II}(r) \to 0\,,\qquad h_{\rm IV}(r) \to 0\,.
\end{equation}
However, since we impose continuity on all metric functions, the ones in
the second de-Sitter shell and between the thin matter shells, have to
approach zero at their outer edges, \ie
\begin{equation}
  h_{\rm III}(r_2) = 1 - \tilde{H}_0^2 r_2^2 \to 0\,,\qquad
  h_{\rm III}(r_3) = 1 - \tilde{H}_0^2 r_3^2 \to 0\,,
\end{equation}
or, equivalently,
\begin{equation}
  \label{eq:r2_r3}
  \tilde{H}_0^2 \to \frac{1}{r_2^2}\,,\qquad \tilde{H}_0^2 \to \frac{1}{r_3^2}\,.
\end{equation}
Bearing in mind that $\tilde{H}_0^2$ is the same constant in region III,
which is a de-Sitter shell, means that fulfilling the
condition~\eqref{eq:r2_r3} is possible only if
\begin{equation}
\label{eq:r3_to_r2}
  r_2 \to r_3\,,
\end{equation}
that is, if the second de-Sitter shell (or region III), has zero radial
extent. However, removing such a shell implies that the two matter shells
II and IV are now adjacent and hence they are effectively a single shell
ranging from $r_2$ to $r_4$, that is, a standard gravastar.
Stated differently, it is not possible to build a meaningful nestar
solution so long as one employs an isotropic pressure, as in the
original Mazur-Mottola thin-shell gravastar\footnote{This result was
already derived in a private discussion with E. Mottola.}. In turn, this
implies that nestars will have to be built with anisotropic pressures, as
we illustrate next.

%=========================================================================
\subsection{Anisotropic-pressure nestar}
\label{sec:nestar}
%=========================================================================

Having shown that an isotropic-pressure (thin-shell) nestar necessarily
coincides with an isotropic-pressure (thin-shell) gravastar, we resort to
the same expedient adopted for gravastars and introduce an anisotropic
pressure. In essence, we use the anisotropic-pressure gravastar from
Sec.~\ref{sec:aniso} and use
Eqs.~\eqref{eq:aniso:line_element}--\eqref{eq:aniso:mass_fun_def}, with
modified continuity conditions~\eqref{eq:aniso:en_den_conds_1} and
\eqref{eq:aniso:en_den_conds_2} to meet the requirements of a nestar
\begin{gather}
  \begin{aligned}\label{eq:nestar:continuity_conditions}
    \text{II.~1st~matter~shell:~}&
    \begin{cases}
      \begin{aligned}
        e(0)    &= e(r_1) = e_1\,,\\[1em]
        e'(r_1) &=e'(r_2)=0\,,\\[1em]
        e''(r_1)&=e''(r_2)=0\,,\\[1em]
        e(r_2)  &= e_3\,,
      \end{aligned}
    \end{cases}\hspace{1em}\\[1.0em]
    \text{IV.~2nd~matter~shell:~}&
    \begin{cases}
      \begin{aligned}
        e(r_2) &= e(r_3) = e_3\,,\\[1em]
        e'(r_3)&=e'(r_4)=0\,,\\[1em]
        e''(r_3)&=e''(r_4)=0\,,\\[1em]
        e(r_4) &= 0\,.    
      \end{aligned}
    \end{cases}
  \end{aligned}
\end{gather}
We can satisfy these conditions via the following layered prescription
for the energy density
\begin{equation}
\hspace{-3.15em}
    e(r) = \left\{\begin{aligned}
    &e_1 = {\rm const.} & \quad {\rm for} \quad & 0\leq r < r_1\,,
    \quad&&\text{:~I.~1st~de-Sitter~shell}\,,\\
    &\sum^{5}_{i=0} a_i\,r^i& \quad {\rm for} \quad & r_1 \leq r \leq r_2\,,
    \quad&&\text{:~II.~1st~matter~shell}\,,\\
    &e_3 = {\rm const.} & \quad {\rm for} \quad & r_2 < r < r_3\,,
    \quad&&\text{:~III.~2nd~de-Sitter~shell}\,, \\
    &\sum^{5}_{i=0} b_i\,r^i&\quad {\rm for} \quad & r_3 \leq r \leq r_4\,,
    \quad&&\text{:~IV.~2nd~matter~shell}\,, \\
    &\quad 0 &\quad {\rm for} \quad & r_4 < r\,,
    \quad&&\text{:~V.~exterior}\,.
    \end{aligned}
    \right.\label{eq:nestar:energy_density}
\end{equation}
Note that modifications to the EOS in the matter shell are also necessary
since Eq.~\eqref{eq:aniso:eos} can only connect an adjacent nonzero
energy-density de-Sitter shell with an adjacent vacuum exterior, while
we here need to connect two nonzero energy-density de-Sitter
shells. As a result, a possible option for the EOSs in the regions I-IV
is as follows
\begin{equation}
\hspace{-3.55em}
    p_r(e) = \left\{\begin{aligned}
    &-e_1& \quad {\rm for} \quad & 0\leq r < r_1\,,\\[0.5em]
    &e\Biggl[\alpha_2-(1+\alpha_2)\frac{(e-e_3)^3-(e-e_1)^3}{(e_1-e_3)^3}\Biggr]&
    \quad {\rm for} \quad & r_1 \leq r \leq r_2\,,\\[0.5em]
    &-e_3& \quad {\rm for} \quad & r_2 < r < r_3\,,\\[0.5em]
    & \left(\frac{e^2}{e_3}\right)
    \Biggl[\alpha_4-(1+\alpha_4)\left(\frac{e}{e_3}\right)^2\Biggr]&
    \quad {\rm for} \quad & r_3 \leq r \leq r_4\,,\\[0.5em]
    &\quad 0 & \quad {\rm for}\quad & r_4 < r\,,
    \end{aligned}
    \right.\label{eq:nestar:eos}
\end{equation}
where the constants $\alpha_2$ and $\alpha_4$ are not constrained yet,
but play the same role of the constant $\alpha$ in the
EOS~\eqref{eq:aniso:eos}. In particular, we can require that the newly
proposed EOS continues to satisfy the following set of
conditions~\cite{Mbonye2005}
\begin{enumerate}
  \label{conditions:nestar_eos}
\item it must satisfy at least the less stringent energy conditions,
  namely, the DEC $(e \geq \lvert p_r \rvert)$ and the
  WEC\footnote{We note that the strong energy condition (SEC)
    requires instead that $e \geq 0$ and $e + p_r + 2 p_t \geq 0$; this
    condition is clearly not satisfied by gravastars nor by our
    nestars. We expect, but do not prove, that when using a different
    prescription of the EOS it should be possible to guarantee that also
    the additional WEC $e + p_t \geq 0$ is met, in contrast to the
    additional DEC $e\geq\lvert p_t \rvert$, which cannot be satisfied by
    ultra-compact objects close to forming an event
    horizon~\cite{Cattoen2005}.}  $(e \geq 0$ and $e + p_r \geq
  0)$~\cite{Mbonye2005};
  \label{item:eos_req3}
\item the corresponding sound speed must always be subliminal;
  \label{item:eos_req2}
\item the corresponding sound speed does not have a maximum at $e=0$.
  \label{item:eos_req1}
\end{enumerate}
Requiring these conditions to hold yields the parameters
\begin{eqnarray}
    \label{eq:nestar_eos_alpha_params}
    &\alpha_2 =\frac{e_1^2 -3e_1e_3+e_3^2}{e_1^2
      -e_1e_3+e_3^2}\,,\\[0.5em]
    &\alpha_4 = \frac{3}{4}
  \Biggl[\sqrt[3]{4-2\sqrt{2}}+\sqrt[3]{2(2+\sqrt{2})}\,\Biggr] 
    \simeq 2.2135\,.
\end{eqnarray}

It is important to underline that within this framework it is not
possible to choose $e_1 = e_3$, as all the polynomial coefficients for
the energy density in region II would reduce to zero except for the first
one, which would be trivially be given by $a_0=e_1=e_3$, thus falling
back into the case of an anisotropic-pressure gravastar. To address this
issue, we impose an additional condition in region II by requiring that
the energy density in the midpoint of region II is proportional to the
energy density in region I, so that the polynomial has to cross at least
one point and does not reduce to a constant, \ie
\begin{equation}
    \label{eq:nestar:cr_k_factor}
  e|_{(r_1+r_2)/2} = (1 - \chi)\,e_1\,,\qquad {\rm with} \qquad \chi \in
    (0,1)\,.
\end{equation}
We note that because of the additional
condition~\eqref{eq:nestar:cr_k_factor}, the polynomial description of
the energy density becomes of sixth order with an additional polynomial
coefficient $a_6$. Furthermore, the introduction of the free parameter
$\chi\in (0,1)$ is useful not only for two nested gravastars, but also when
we will consider $N$ nested gravastars in Sec.~\ref{sec:ml_nestar}.

Given the generality of the energy density and EOS considered, there is a
large freedom in the construction of two nested gravastars and in what
follows we consider two specific classes. The first one is reminiscent of
a ``copy-paste'' operation, in that the energy density of regions I and
III are set to be the same, namely, $e_1=e_3$. In this case, because the
EOS~\eqref{eq:nestar:eos} in region II would have a zero in the
denominator, we need to use in such a region the same EOS used for a
single gravastar~\eqref{eq:aniso:eos}, as it is capable of connecting two
de-Sitter shells with the same energy density.

\begin{figure*}
  \centering
  \hspace{1.0em}
  \includegraphics[width=0.45\textwidth]{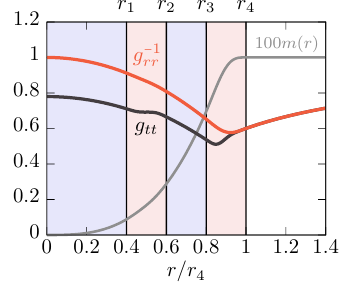}
  \includegraphics[width=0.45\textwidth]{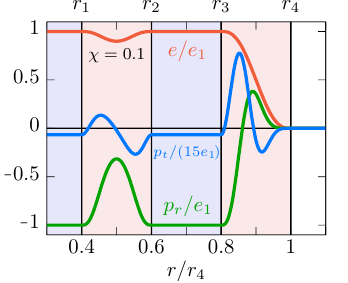}
  \caption{The same as in
    Fig.~\ref{fig:anisotropic:en_pres_metr_mass} but for a
    representative nestar with a mass $M$, outer radius
    $r_4=5\,M$, and where energy densities in regions I and III are the
    same, \ie $e_1=e_3$ and $\chi=0.1$
    from~\eqref{eq:nestar:cr_k_factor}. For better visualisation, the
    tangential pressure is scaled by a factor of $1/15$, while the mass
    function is scaled by a factor of $100$.}
\label{fig:nestar:en_pres_metr_mass_eqe}
\end{figure*}

A bit of algebra allows one to deduce that the set of unknown
coefficients in the energy-density function for the first matter shell in
region II are given by
\begin{align}
  \hspace{-2.05cm}
  \begin{aligned} 
    a_6&=\frac{64\chi\,e_1}{(r_2-r_1)^6}\,,& \quad a_5&=-\frac{192\chi\,e_1(r_1+r_2)}{(r_2-r_1)^6}\,,\nonumber\\[1em]
    a_4&= \frac{192\chi\,e_1(r_1^2+3r_1r_2+r_2^2)}{(r_2-r_1)^6}\,, & \quad 
    a_3&=-\frac{64\chi\,e_1(r_1+r_2)
      (r_1^2+8r_1r_2+r_2^2)}{(r_2-r_1)^6}\,,\\[1em]
    a_2&= \frac{192
      \chi\,e_1  r_1r_2 (r_1^2+3r_1r_2+r_2^2)}{(r_2-r_1)^6}\,,& 
    a_1&=- \frac{192 \chi\,e_1 r_1^2r_2^2 (r_1+r_2)}{(r_2-r_1)^6}\,,\nonumber\\[1em]
    a_0&= e_1  +\frac{64\chi\,e_1 r_{1}^3r_{2}^3 }{(r_{2}-r_{1})^6}\,,\\
  \end{aligned}
  \\
\end{align}
while the coefficients in the energy-density function for the second matter shell
in region IV are given by
\begin{align}
\hspace{-2.4em}
\begin{aligned}
  b_5&=-\frac{6e_1}{(r_4-r_3)^5}\,, & \qquad \hspace{0.44em} b_4&=\frac{15e_1(r_3+r_4)}{(r_4-r_3)^5}\,,\nonumber\\[1em]
  b_3&= -\frac{10e_1(r_3^2+4r_3r_4+r_4^2)}{(r_4-r_3)^5}\,, & \qquad b_2&=\frac{30  e_1 r_3 r_4 (r_3+r_4)}{(r_4-r_3)^5}\,,\\[1em]
  b_1&=- \frac{30 e_1 r_3^2r_4^2}{(r_4-r_3)^5}\,, &\qquad b_0&=\frac{
    e_1 r_4^3(10r_3^2-5r_3r_4+r_4^2)}{(r_4-r_3)^5}\,. 
\end{aligned}
\\
\end{align}

Figure~\ref{fig:nestar:en_pres_metr_mass_eqe} shows a representative
solution for the case of two nested gravastars having the same energy
density in the de-Sitter shells. In particular, it refers to a nestar
with a mass $M$ and outer radius $r_4=5\,M$. Note that all the metric
functions and fluid variables, as well as their first derivatives, are
continuous; the tangential
pressure~\eqref{eq:anisotropic_tangential_pressure} is still the dominant
one and scaled by a factor of $1/15$, while the mass function is scaled
by a factor of $100$ for better visibility.
 
Following a similar logic, it is possible to construct equilibrium
solutions also for two nested gravastars when the de-Sitter shells have
different energy densities $e_1\neq e_3$. Also in this case,
straightforward algebra allows one to deduce that the set of unknown
coefficients in the energy-density function for the first matter shell in
region II are given by
\begin{align}
  \hspace{-5.5em}
  \begin{aligned}
    a_5&=\frac{6(e_3-e_1)}{(r_2-r_1)^5}\,,& \,
    a_4&=-\frac{15 (e_3-e_1)(r_1+r_2)}{(r_2-r_1)^5}\,,\nonumber\\[1em]
    a_3&= \frac{10 (e_3-e_1)(r_1^2+4r_1r_2+r_2^2)}{(r_2-r_1)^5}\,,& \,
    a_2&=-\frac{30 (e_3-e_1) (r_1+r_2)r_1 r_2}{(r_2-r_1)^5}\,,\\[1em]
    a_1&= \frac{30 (e_3-e_1)r_1^2r_2^2 }{(r_2-r_1)^5}\,,&\,
    a_0&=e_3 - \frac{(e_3-e_1) (10r_1^2-5r_1r_2+r_2^2)r_2^3}{(r_2-r_1)^5}\,,\nonumber
  \end{aligned}
  \\
\end{align}
while the coefficients in the energy-density function for the second matter shell
in region IV are given by
\begin{align}
  \hspace{-2.4em}
  \begin{aligned}
    b_5&=-\frac{6e_3}{(r_4-r_3)^5}\,,& \qquad \hspace{0.44em}
    b_4&=\frac{15e_3(r_3+r_4)}{(r_4-r_3)^5}\,,\nonumber\\[1em]
    b_3&= -\frac{10e_3(r_3^2+4r_3r_4+r_4^2)}{(r_4-r_3)^5}\,,& \qquad
    b_2&=\frac{30 e_3 r_3 r_4 (r_3+r_4)}{(r_4-r_3)^5}\,,\\[1em]
    b_1&=- \frac{30 e_3 r_3^2r_4^2}{(r_4-r_3)^5}\,,&\qquad b_0&=\frac{e_3
      r_4^3 (10r_3^2-5r_3r_4+r_4^2)}{(r_4-r_3)^5}\,. 
  \end{aligned}
  \\
\end{align}

\begin{figure*}
  \centering
  \hspace{1.0em}
  \includegraphics[width=0.45\textwidth]{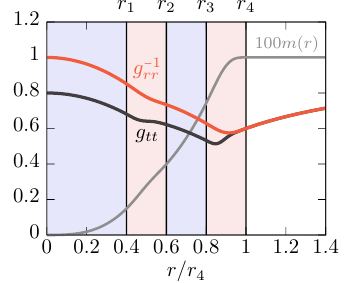}
  \includegraphics[width=0.45\textwidth]{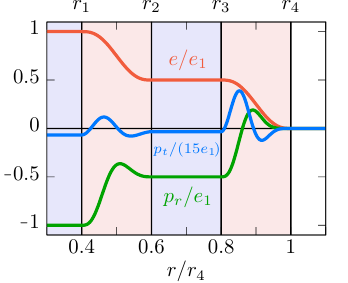}\\[1em]
  \hspace{1.0em}
  \includegraphics[width=0.45\textwidth]{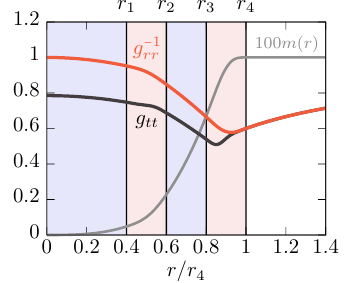}
  \includegraphics[width=0.45\textwidth]{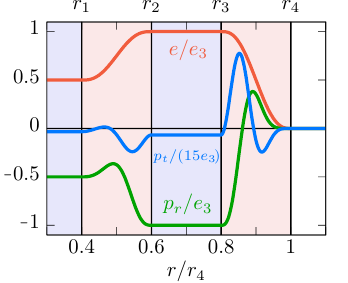}
\caption{The same as in
  Fig.~\ref{fig:nestar:en_pres_metr_mass_eqe} but for a
  representative nestar with different energy densities in the matter
  shells. The upper row refers to a nestar with decreasing energy
  densities $e_1>e_3$, while the lower row to one with increasing energy
  densities $e_1<e_3$.}
\label{fig:nestar:en_pres_metr_mass}
\end{figure*}

Figure~\ref{fig:nestar:en_pres_metr_mass} provides the equivalent
information as Fig.~\ref{fig:nestar:en_pres_metr_mass_eqe} but for a
nestar having de-Sitter shells with decreasing energy density $e_1>e_3$
(upper row) and an increasing one $e_1<e_3$ (lower row). Note that also
in these cases, all the metric functions and fluid variables, as well as
their first derivatives, are continuous. Similarly, the tangential
pressure still provides the dominant contribution to the equilibrium;
note that the models presented in Figs.~\ref{fig:nestar:en_pres_metr_mass_eqe}
and~\ref{fig:nestar:en_pres_metr_mass}, but also those that we will
discuss in the following figures, do satisfy the DEC and WEC.

%=========================================================================
\subsection{Space of solutions of anisotropic-pressure nestars}
\label{sec:space_sol_nestar}
%=========================================================================

The systematic analysis we have carried out for the anisotropic-pressure
gravastar can be extended in the case of a nestar, where all possible
equilibrium configurations need to respect the no-horizon condition
\eqref{eq:aniso:no_horizon_inequality}. The main difference here is that
the space of solutions will obviously be larger and can be characterised
in terms of the thickness of the two matter shells, namely,
$\delta_2:=r_2-r_1$ and $\delta_4:=r_4-r_3$, and of the overall
compactness $\mathcal{C}:= M/r_4$.

\begin{figure*}
  \centering
  \hspace{2.0em}
  \includegraphics[width=0.75\textwidth]{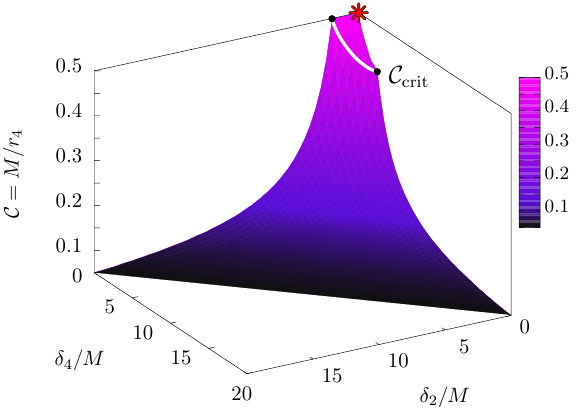}
  \caption{Space of possible solutions for nestars represented in terms
    of the nestar compactness $\mathcal{C}:= M/r_4$ and of the
    thicknesses of the matter shells $\delta_2:=r_2-r_1$,
    $\delta_4:=r_4-r_3$. The coloured surface marks the limit of the
    space of solutions beyond which an event horizon appears or $r_1<0$,
    while the white line shows the position of the critical compactness
    $\mathcal{C}_{\rm crit}$ dividing the two branches already encountered
    for a single-shell gravastar. Shown with a red asterisk is the
    isotropic-pressure (thin-shell) gravastar solution.}
  \label{fig:nestar_sol_space}
\end{figure*}

Figure~\ref{fig:nestar_sol_space} shows the space of all such possible
solutions of the nestar compactness $\mathcal{C}$, when expressed in
terms of two matter-shell thicknesses $\delta_2$ and $\delta_4$. As for
gravastars, here the two-dimensional surface marks the limit above which
no nestar solution can be found because the no-horizon condition does not
hold anymore or the matter shells already encompass the whole
nestar. Note that the overall behaviour is similar but also distinct from
that already encountered for the thin-shell gravastar in
Fig.~\ref{fig:MM:gravastar} and for the anisotropic-pressure gravastar
in Fig.~\ref{fig:anis:sol_space}.

In particular along the surface, for specific combinations of $\delta_2$
and $\delta_4$, there exists a critical compactness $\mathcal{C}_{\rm
  crit}$ distinguishing the restricted and unrestricted branch of the
space of solutions and this is marked with a white solid line on the
limiting surface in Fig.~\ref{fig:nestar_sol_space}. Solutions
approaching the surface beneath the critical line can be described with a
generalisation of the hyperbolic condition for two matter shells
\begin{equation}
  \label{eq:nestar:hyperbolic_3d}
  \mathcal{C}\,\frac{\delta_2+\delta_4}{M} =
  \frac{\delta_2+\delta_4}{r_4}\to 1 \qquad \Rightarrow \qquad
  \delta_2+\delta_4
  \to r_4\,,
\end{equation}
which can be interpreted as the case in which both matter shells
encompass the whole nestar, thus in full analogy with what we have seen
in the case of the gravastar, where the hyperbolic
condition~\eqref{eq:hyperb_cond} also referred to the situation in which
the matter shell encompassed the whole object. Therefore, the surface
below the critical line marks the unrestricted solutions, while the red
asterisk helps locate the position in the space of solutions of the
isotropic-pressure (thin-shell) gravastar.

\begin{figure*}
  \centering\hspace{1.0em}
  \includegraphics[width=0.45\textwidth]{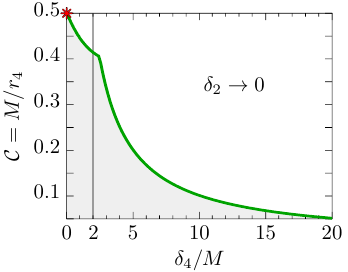}
  \includegraphics[width=0.45\textwidth]{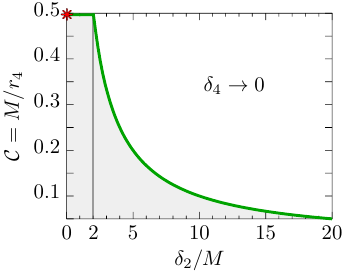}
  \caption{Sections at constant values of the inner matter shell
    thickness ($\delta_2\to0$, left panel) or at constant outer matter
    shell thickness ($\delta_4\to0$, right panel) for the compactness of
    the admitted solutions for nestars. The left panel coincides with the
    limit of a standard anisotropic-pressure gravastar, while the right
    panel reports the disappearance of the restricted branch and thus
    yields solutions approaching a compactness of $\mathcal{C}\to(1/2)^{-}$
    whilst being comprised of essentially only matter. Shown with a red
    asterisk is the isotropic-pressure (thin-shell) gravastar solution.}
  \label{fig:nestar_sol_space_sec}
\end{figure*}

Shown instead in Fig.~\ref{fig:nestar_sol_space_sec} are two sections of
the space of solutions for the compactness of the nestar at constant
values of the inner-shell thickness ($\delta_2 \to 0$, left panel) or at
constant outer-shell thickness ($\delta_4 \to 0$, right panel) for the
compactness. Starting from the left panel, it is easy to realise that
this limit corresponds essentially to a standard anisotropic-pressure
gravastar [\cf Fig.~\ref{fig:anis:sol_space}]; indeed, $\delta_4$
exhibits a critical thickness which is the same encountered for an
anisotropic-pressure gravastar, \ie $\delta_{\rm crit} := (\delta_{2,
  {\rm crit}},\delta_{4,{\rm crit}}) \simeq (0,2.4740\,M)$ [also in this
  case, we can mark a critical compactness $\mathcal{C}_{\rm crit}
  \simeq0.4042$]. Far more interesting is the space of solutions shown in
the right panel (\ie for $\delta_4 \to0$) for in this case the restricted
branch disappears, and where nestar solutions approaching the green line
have a particularly interesting structure. More specifically, the outer
de-Sitter and the outer matter shell (regions III and IV) are of
infinitesimal size ($r_2\to r_3 \to r_4 > 2\,M$) because the green line
corresponds to $\mathcal{C}\delta_2/M\to1\Rightarrow\delta_2\to r_4$.  As
a result, the inner de-Sitter shell (region I) is also infinitesimal
($r_1\to 0$) and the inner matter shell (region II) fills the whole
nestar. In this way, the critical thickness is moved to $\delta_{\rm
  crit} \to (2\,M^{+},0)$ and the critical compactness to $\mathcal{C}
_{\rm crit} \to (1/2)^{-}$; clearly, the condition $r_2 \to r_3 \to r_4
\to 2\,M^{+}$ is met when the compactness is the maximal allowed, \ie
$\mathcal{C} \to (1/2)^{-}$. This is a particularly interesting
configuration, where the compactness is maximal but the interior of the
compact object is essentially all filled of matter. This is to be
contrasted with the isotropic-pressure, thin-shell gravastar model that
has $r_1 \to r_2 \to 2M$ and hence where the compact object is
essentially all filled by the de-Sitter portion. A more detailed
discussion of the radial structure of the nestar solutions along the two
sections $\delta_2 \to 0$ and $\delta_4 \to 0$ can be found in
Appendix~\ref{appendix:radstruc}.

As a concluding remark, we note that what makes the nestar solution with
$r_1\to 0$, $r_2 \to r_3 \to r_4 \to 2\,M^{+}$, and $\mathcal{C} \to
(1/2)^{-}$ particularly fascinating is that it effectively replaces a
Schwarzschild black-hole with a solution that has essentially the same
compactness, but where the singularity is replaced by a regular
point-like de-Sitter interior and where the null surface is replaced by
an ordinary time-like surface with two infinitesimal shells (de-Sitter
and matter), while the rest is filled with matter satisfying the WEC and
DEC. Hence, when going back to the issue of the genesis of these
speculative objects, the nestar in question could be considered as a more
plausible endpoint of a gravitational collapse since the appearance of a
singularity would be replaced by the appearance of only a point-like
de-Sitter mini-universe and not a whole de-Sitter interior.

%=========================================================================
\subsection{Anisotropic-pressure multi-layer nestar}
\label{sec:ml_nestar}
%=========================================================================

We conclude our discussion by generalising many of the concepts and the
formalism introduced so far in the case of two nested gravastars to the
case of $K$ nested gravastars, where $K$ is an arbitrary integer. Once
again, intuition serves as a useful guide to understand why a
construction with $K$ nested gravastars should be reasonable if an
equilibrium solution can be found for two. Let us recall that because of
the balanced interplay between the expanding interior and the contracting
shell, each gravastar is essentially in equilibrium, so that the
operation of nesting gravastars into each other will not radically change 
such an equilibrium. Exactly for this reason, it is reasonable to expect
that the same logic should apply also when the nesting is done for two
gravastars or for $K$, with $K$ arbitrary.

In practice, we proceed exactly as done in the case of \textit{two}
nested gravastars in Sec.~\ref{sec:nestars} [see
  Eqs.~\eqref{eq:N:eos1}--\eqref{eq:N:eos5}] , and split now the
spacetime not into five regions but into $2K+1$ regions, $K$ of which are
filled by de-Sitter shells with constant energy densities and negative
pressures, $K$ by matter shells with however complicated EOSs, and the
$2K+1$-th region is that with a vacuum and hence described by the
Schwarzschild solution. In essence, we describe the nesting of $K$
gravastars to produce a $2K$-shells nestar using the following scheme
\begin{eqnarray}
  &
  \text{de-Sitter~shells}  \hspace{2.0cm}
  &r \in\left(r_{2k-2}, r_{2k-1}\right)\,,\\
  &
  \text{matter~shells}
  &r \in\left[r_{2k-1}, r_{2k}\right]\,,\\
  &
  \text{exterior}
  &r \in\left(r_{2K}, \infty\right)\,, \\ 
  &&\hspace{3.0cm} k\in\mathbb{N}\,, \quad k\in [1,K]\,, \nonumber 
\end{eqnarray}
where the radii $r_{2k-1}$ and $r_{2k}$ are free parameters except for
the centre, which obviously needs to be at $r_0=0$; the edge of the
outermost shell, and thus of the nestar, is at $r_{2K}$.

Recalling that the (odd) interior de-Sitter shells are of constant energy
density while the (even) matter-shell regions have energy densities
expressed in terms of polynomial functions, the complete energy-density
function can be written in a compact manner as
\begin{align}
e(r) = \left\{
\begin{aligned}
  &e_{2k-1}={\rm const.}    & \quad {\rm for} \quad &r \in\left(r_{2k-2}, r_{2k-1}\right)\,,\\
  &E_{2k}(r)                & \quad {\rm for} \quad &r \in\left[r_{2k-1}, r_{2k}\right]\,,\\
  &e_{2K+1}=0               & \quad {\rm for} \quad &r \in\left(r_{2K}, \infty\right)\,, 
\end{aligned}
\right.
\label{eq:k-nestar:energy_density}\\
&&&\hspace{-1.29cm} k\in\mathbb{N}\,, \quad k\in [1,K]\,,\nonumber
\end{align}
where $e_{2k-1}$ are free parameters. Depending on whether the energy
density has to connect two de-Sitter shells with the same energy density
or two different ones, one will either choose a 6th-order or a 5th-order
polynomial ($e_{2K+1}=0$ so that the last shell is 5th-order), namely
\begin{align}
  E_{2k}(r) = \left\{\begin{aligned}
  &\sum^{6}_{i=0}        a_{i,2k}\,r^i\,,  &\quad  & e_{2k-1}=e_{2k+1}\,,\\
  &\sum^{5}_{i=0} \tilde{a}_{i,2k}\,r^i\,, &\quad  & e_{2k-1}\neq e_{2k+1}\,,
  \end{aligned}
  \right.\\
&&&\hspace{+0.09cm} k\in\mathbb{N}\,, \quad k\in [1,K]\,;\nonumber
\end{align}
the full expressions for the coefficients ${a}_{i,2k}$ and
$\tilde{a}_{i,2k}$ as a function of the various radii and energy
densities can be found in Appendix~\ref{appendix:nestar_coefficients}.

Similarly, the radial pressure can be expressed as
\begin{align}
p_r(e) = \left\{
\begin{aligned}
  &-e_{2k-1}  & \quad {\rm for} \quad &r \in\left(r_{2k-2}, r_{2k-1}\right)\,,\\
  &P_{2k}(e)  & \quad {\rm for} \quad &r \in\left[r_{2k-1}, r_{2k}\right]\,,\\
  &\quad 0   & \quad {\rm for} \quad &r \in\left(r_{2K}, \infty\right)\,, 
\end{aligned}
\right.
\label{eq:k-nestar:pressure}\\
&&&\hspace{-0.51cm} k\in\mathbb{N}\,, \quad k\in [1,K]\,,\nonumber
\end{align}
where the de-Sitter shells have a simple $p=-e$ EOS in the spirit of a
de-Sitter spacetime. The EOSs for the matter-shell regions are instead
computed depending on whether the energy densities are the same or
different across the adjacent de-Sitter shell regions\footnote{It is
possible to choose between both EOSs for the last shell because the EOS
for $e_{2k-1}=e_{2k+1}$ is capable of connecting a de-Sitter
shell and a vacuum.}
\begin{align}
\hspace{-2.46cm}
  P_{2k}(e) = \left\{\begin{aligned}
  &\frac{E_{2k}^2(r)}{e_{2k-1}}\Biggl[\alpha_{2k}-(1+\alpha_{2k})
    \left(\frac{E_{2k}(r)}{e_{2k-1}}\right)^2\Biggr]\,,
  \hspace{2.06cm} e_{2k-1}=e_{2k+1}\,,\\[0.7em]
  &E_{2k}(r)\Biggl[\alpha_{2k}-(1+\alpha_{2k})\frac{(E_{2k}(r)-e_{2k+1})^3 -
      (E_{2k}(r)-e_{2k-1})^3}{(e_{2k-1}-e_{2k+1})^3}\Biggr]\,, \\[0.5em]
  &\hspace{8.0cm}  e_{2k-1}\neq e_{2k+1}\,,
  \end{aligned}
  \right.\\[1.0em]
&&&\hspace{-2.88cm} k\in\mathbb{N}\,, \quad k\in [1,K]\,,\nonumber
\end{align}
modifying the definitions of~\eqref{eq:nestar_eos_alpha_params} for a
$2K$-layers nestar
\begin{align}
  \hspace{-3.5em}\alpha_{2k} = \left\{\begin{aligned}
  &\frac{3}{4}\Biggl[\sqrt[3]{4-2\sqrt{2}}+\sqrt[3]{2(2+\sqrt{2})}\,\Biggr] 
\simeq 2.2135\,,  &  e_{2k-1}=e_{2k+1}\,,\\[0.5em]
  &\frac{e_{2k-1}^2
      -3e_{2k-1}e_{2k+1}+e_{2k+1}^2}{e_{2k-1}^2
      -e_{2k-1}e_{2k+1}+e_{2k+1}^2}\,,   &  e_{2k-1}\neq e_{2k+1}\,,
      \label{eq:nestar_inf_eos_alpha_params}
      \end{aligned}
  \right.\\
&&&\hspace{-3.53cm}\hspace{3.5em} k\in\mathbb{N}\,, \quad k\in [1,K]\,.\nonumber
\end{align}
Finally, the tangential pressure can be computed using
Eq.~\eqref{eq:anisotropic_tangential_pressure}, so that, after choosing
the radii $r_{2k-1}$, $r_{2k}$ and the de-Sitter energy densities
$e_{2k-1}$, it is possible to build a $2K$-layers nestar with Eqs.~
\eqref{eq:k-nestar:energy_density}--\eqref{eq:nestar_inf_eos_alpha_params}
and the polynomial coefficients that are reported in
Appendix~\ref{appendix:nestar_coefficients}.

We conclude this section and this work by pushing even further the level
of speculation. Given that: (i) a nestar with an arbitrarily large number
of layers is possible to construct; (ii) that the maximum compactness
that is possible to reach is smaller but arbitrarily close to that of a
Schwarzschild black hole; (iii) the volume fraction of the nestar that
satisfies the WEC and DEC can be taken arbitrarily close to one, a
multi-layer nestar could represent an interesting alternative to a black
hole solution with the advantage of being everywhere regular and
nonvacuum.

%-------------------------------------------------------------------------
\section{Conclusion}
\label{sec:conclusion}
%-------------------------------------------------------------------------

More than 20 years ago, Mazur and Mottola proposed the isotropic
(thin-shell) gravastar model~\cite{Mazur2001, Mazur2004}, which has since
then attracted attention and triggered considerable related work. More
importantly, the gravastar model has suggested a new way of thinking
about compact objects and whether alternatives to black holes without an
event horizon can be envisaged as the final point of gravitational
collapse. Twenty years later, and while many of the aspects of these
objects are now well understood, a number of issues remain unsolved about
their dynamical formation or the corresponding
observability~\cite{Broderick2007, Carballo2018, EHT_SgrA_PaperVI_etal,
  Carballo2022}.

We have here revisited the gravastar model and raised the bar of
speculation by considering new solutions that are inspired by the
original model but also offer surprising new features. In particular, we
have shown that when using anisotropic pressure it is possible to nest
two gravastars into each other and obtain a new and rich class of
solutions of the Einstein equations. Indeed, thanks to the fact that each
gravastar essentially behaves as a distinct self-gravitating equilibrium
configuration, a very large space of parameters exists for nested
gravastars having an anisotropic pressure support. In this space of
solutions, one is particularly interesting since it replaces a
Schwarzschild black-hole with a solution that has essentially the same
compactness, but where the singularity is replaced by a regular
point-like de-Sitter interior and where the null surface is replaced by
an ordinary time-like surface with two infinitesimal shells (de-Sitter
and matter), while the rest is filled with matter satisfying the WEC and
DEC. Finally, we have shown that the nested two-gravastar solutions can
be extended to an arbitrarily large number of elements, with a
prescription that can be easily specified in terms of recursive
relations.

Although these equilibria are admittedly on the exotic side of
astrophysical compact objects, it is interesting to note that more than
100 years after the introduction of general relativity, new solutions can
be still found in one of the simplest spacetimes: spherically symmetric,
static, and not in vacuum. The results presented here can be further
explored in a number of different directions, which include: the study of
different EOSs, the stability of these solutions against perturbations
and the corresponding quasi-normal modes, the existence of stable and
unstable circular photon orbits (see~\cite{Carballo2023} for a
comprehensive discussion), the extension to rotating configurations (the
different shells can have the same or different angular velocities) and
their stability~\cite{Chirenti2008}, and, of course, the exploration of
similar equilibria in other theories of gravity. Among these possible
developments, the exploration of the response of nestars to
perturbations probably represents the most interesting avenue, as it
will offer, besides a confirmation about the stability of these compact
objects, also precise tools to distinguish nestars from gravastars and
hence from black holes.

\begin{figure}
  \centering\hspace{-0.5em}
  \includegraphics[width=0.45\textwidth]{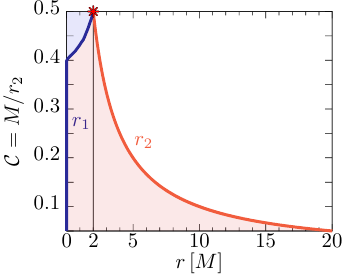}
  \caption{Radial structure of anisotropic-pressure gravastars on the
    outer edge of the space of possible solutions (\ie along the green
    line in Fig.~\ref{fig:anis:sol_space}). Shown with a red asterisk is
    the isotropic-pressure (thin-shell) gravastar solution.}
  \label{fig:radstruc_gravastar}
\end{figure}    

%-------------------------------------------------------------------------
\section{Appendix}
\label{sec:appendix}
%-------------------------------------------------------------------------

%=========================================================================
\subsection*{Radial structure of anisotropic-pressure gravastars and nestars}
\label{appendix:radstruc}
%=========================================================================

As an example of the complex radial structure of anisotropic-pressure
gravastars, we report in Fig.~\ref{fig:radstruc_gravastar} the values of
the radii $r_1$ and $r_2$ as a function of the compactness
$\mathcal{C}=M/r_2$ for a sequence of gravastars on the outer edge of the
space of possible solutions (\ie along the green line in
Fig.~\ref{fig:anis:sol_space}). Note how for compactnesses smaller than
the critical compactness $\mathcal{C}_{\rm crit} \simeq 0.4042$, the
solutions are those of the unrestricted branch and hence with an
infinitesimal de-Sitter interior ($r_1 \to 0$) and a matter shell that
fills the whole gravastar. However, for compactnesses larger than the
critical one, the space of solutions jumps to the restricted branch and
the de-Sitter interior expands. When $\mathcal{C} \to (1/2)^{-}$, the
matter shell shrinks to an infinitesimal thickness ($r_1 \to 2\,M^{-}$
and $r_2 \to 2\,M^{+}$) and the de-Sitter interior tends to fill the
entire gravastar.

\begin{figure*}
  \centering
  \includegraphics[width=0.45\textwidth]{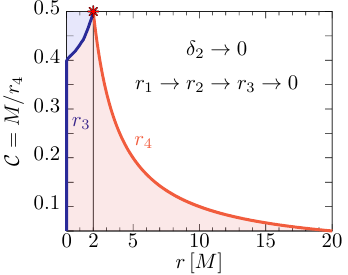}
  \includegraphics[width=0.45\textwidth]{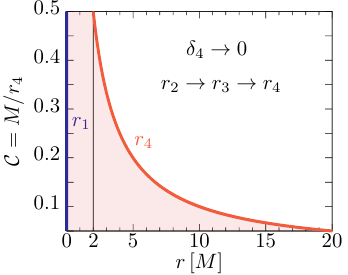}
  \caption{Radial structure for a sequence of $K=2$ nestars on the outer
    edge of the space of possible solutions, \ie along the green lines in
    the left and right panels of Fig.~\ref{fig:nestar_sol_space_sec}. The
    left panel refers to a section at constant value of the inner-shell
    thickness ($\delta_2 \to 0$), while the right panel to a constant
    outer-shell thickness ($\delta_4 \to 0$). Shown with a red asterisk
    is the isotropic-pressure (thin-shell) gravastar solution.}
  \label{fig:radstruc_nestar}
\end{figure*}

Similarly, we show in Fig.~\ref{fig:radstruc_nestar} the values of the
radii $r_1, r_2, r_3$, and $r_4$ for a sequence of $K=2$ nestars on the
outer edge of the space of possible solutions. In particular, the left
panel refers to a section at zero value of the inner matter shell
thickness ($\delta_2 \to 0$), while the right panel to a zero outer
matter shell thickness ($\delta_4 \to 0$). In other words, the left and
right panels show respectively the radial structure of the nestars along
the green lines in the left and right panels of
Fig.~\ref{fig:nestar_sol_space_sec}. Note that the left panel of
Fig.~\ref{fig:radstruc_nestar} is very similar to
Fig.~\ref{fig:radstruc_gravastar}, with the only difference that now
regions I, II and III of the spacetime have infinitesimal thickness ($r_1
\to r_2 \to r_3\to 0$); in this respect, these nestars are effectively
equivalent to gravastars with anisotropic pressure.

On the other hand, the right panel highlights a rather different radial
structure. As already anticipated in the Sec.~\ref{sec:space_sol_nestar},
in this case, all solutions are on the unrestricted branch, so that the
outer de-Sitter and the outer matter shells (regions III and IV) are of
infinitesimal size ($r_2\to r_3 \to r_4 > 2\,M$), because the green line
in the right panel of Fig.~\ref{fig:nestar_sol_space_sec}
corresponds to $\mathcal{C}\delta_2/M\to1\Rightarrow\delta_2\to r_4$.
As a result, the inner de-Sitter shell (region I) is also infinitesimal
($r_1\to 0$) and the inner matter shell (region II) fills the whole nestar.
In this way, the critical thickness is moved to $\delta_{\rm crit} :=
(\delta_{2,{\rm crit}},\delta_{4,{\rm crit}}) \to 
(2\,M^{+},0)$ and the critical compactness to $\mathcal{C}_{\rm crit}
\to (1/2)^{-}$; clearly, the radii must be such that $r_2 \to r_3 \to r_4
\to 2\,M^{+}$ when the compactness reached is the maximum allowed,
\ie $\mathcal{C} \to (1/2)^{-}$. As mentioned in the main text, this is a
particularly interesting configuration, where the compactness is maximal
but the interior of the compact object is essentially all filled of
matter.

\begin{figure*}
  \centering
  \hspace{-0.4em}
  \includegraphics[width=0.31\textwidth]{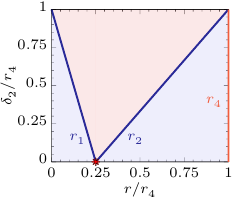}
  \hspace{0.2cm}
  \includegraphics[width=0.31\textwidth]{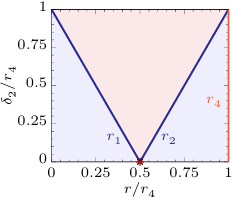}
  \hspace{0.2cm}
  \includegraphics[width=0.31\textwidth]{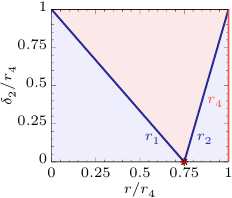}
  \caption{Examples of the radial structure for a sequence of $K=2$
    nestars along $\mathcal{C}={\rm const.}$ lines in the $\delta_4\to0$
    section of Fig.~\ref{fig:nestar_sol_space_sec} with $r_3\to
    r_4$. Shown are different cases, where each de-Sitter shell takes up
    different fractions of the available space, namely $r_1 =
    \psi(1-\delta_2/r_4)$ and $r_2=r_1+\delta_2$, with $\psi=1/4$ in the
    left panel, $\psi=1/2$ in the middle panel and $\psi=3/4$ in the
    right panel. Note that the first matter shell, shown with a red
    shading above, can be placed arbitrarily inside the nestar and does
    not necessarily have to follow the radial structure suggested
    here. Interestingly, when $\mathcal{C}\to(1/2)^{-}$, $\delta_2\to0$
    and $\delta_4\to0$, the solutions correspond to the
    isotropic-pressure (thin-shell) gravastar (see also left panel of
    Fig.~\ref{fig:radstruc_nestar}), which is marked with a red
    asterisk.}
  \label{fig:shell_placement_nestar}
\end{figure*}

Finally, Fig.~\ref{fig:shell_placement_nestar} provides some
representative examples of the radial structure for a sequence of $K=2$
nestars along $\mathcal{C}={\rm const.}$ lines in the $\delta_4\to0$
section of Fig.~\ref{fig:nestar_sol_space_sec} with $r_3\to r_4$. In
particular, setting $r_1=\psi(1-\delta_2/r_4)$ and $r_2=r_1+\delta_2$,
the different cases show examples where each de-Sitter shell takes up
different fractions of the available space, namely with $\psi=1/4$ in the
left panel, $\psi=1/2$ in the middle panel and $\psi=3/4$ in the right
panel.  Note that the first matter shell, shown with a red shading, can
be placed arbitrarily inside the nestar and does not necessarily have to
follow the radial structure suggested here. Particularly interesting are
the cases when $\mathcal{C}\to(1/2)^{-}$, $\delta_2\to0$ and
$\delta_4\to0$, which essentially correspond to the isotropic-pressure
(thin-shell) gravastar (see also left panel of
Fig.~\ref{fig:radstruc_nestar}).

%=========================================================================
\subsection*{N-layer nestar coefficients}
\label{appendix:nestar_coefficients}
%=========================================================================

We here report the polynomial coefficients needed for the energy-density
of an anisotropic-pressure nestar with $2K$ shells. As done for the
simpler nestar with four shells, we need to distinguish the matter shells
regions adjacent to de-Sitter shells having the same energy density from
those where the energy densities are different. In the first case, \ie
for $e_{2k-1} = e_{2k+1}$, the coefficients are found to be given by the
following recursive relations
\begin{align}
\hspace{-5.1em}
\begin{aligned}
  a_{6,2k}&=\frac{64\,\chi_{2k}\,e_{2k-1}}{(r_{2k}-r_{2k-1})^6}\,,\\[1em]
  a_{5,2k}&=-\frac{192\,\chi_{2k}\,e_{2k-1}(r_{2k-1}+r_{2k})}{(r_{2k}-r_{2k-1})^6}\,,\\[1em]
  a_{4,2k}&=
  \frac{192\,\chi_{2k}\,e_{2k-1}(r_{2k-1}^2+3r_{2k-1}r_{2k}+r_{2k}^2)}{(r_{2k}-r_{2k-1})^6}\,,\\[1em]
  a_{3,2k}&=-\frac{64\,\chi_{2k}\,e_{2k-1}(r_{2k-1}+r_{2k})
    (r_{2k-1}^2+8r_{2k-1}r_{2k}+r_{2k}^2)}{(r_{2k}-r_{2k-1})^6}\,,\\[1em]
  a_{2,2k}&= \frac{192\,\chi_{2k}\,e_{2k-1}\,r_{2k-1}\,r_{2k}\,(r_{2k-1}^2+3r_{2k-1}r_{2k}+r_{2k}^2)}{(r_{2k}-r_{2k-1})^6}\,,\\[1em]
  a_{1,2k}&=- \frac{192\,\chi_{2k}\,e_{2k-1}\,r_{2k-1}^2\,r_{2k}^2\,(r_{2k-1}+r_{2k})}{(r_{2k}-r_{2k-1})^6}\,,\\[1em]
  a_{0,2k}&= e_{2k-1}  +\frac{64\,\chi_{2k}\,e_{2k-1}\,r_{2k-1}^3\,r_{2k}^3}{(r_{2k}-r_{2k-1})^6}\,.\\ 
\end{aligned}   
\\\nonumber k\in\mathbb{N}\,, \quad k\in [1,K]\,. \hspace{-2.45cm}
\end{align}

On the other hand, in the second case, \ie for $e_{2k-1} \neq e_{2k+1}$,
the coefficients are given by the recursive relations
\begin{align}
\hspace{-5.1em}
\begin{aligned}
  \tilde{a}_{5,2k}&=\frac{6(e_{2k+1}-e_{2k-1})}{(r_{2k}-r_{2k-1})^5}\,,
  \\[1em]
  \tilde{a}_{4,2k}&=-\frac{15(e_{2k+1}-e_{2k-1})(r_{2k-1}+r_{2k})}{(r_{2k}-r_{2k-1})^5}\,,\\[1em]
  \tilde{a}_{3,2k}&=
  \frac{10(e_{2k+1}-e_{2k-1})(r_{2k-1}^2+4r_{2k-1}r_{2k}+r_{2k}^2)}{(r_{2k}-r_{2k-1})^5}\,,\\[1em]
  \tilde{a}_{2,2k}&=-\frac{30 (e_{2k+1}-e_{2k-1})
    (r_{2k-1}+r_{2k})r_{2k-1} r_{2k}}{(r_{2k}-r_{2k-1})^5}\,, \\[1em]
  \tilde{a}_{1,2k}&= \frac{30 (e_{2k+1}-e_{2k-1}) r_{2k-1}^2r_{2k}^2}{(r_{2k}-r_{2k-1})^5}\,,\\[1em]
  \tilde{a}_{0,2k}&=e_{2k+1} - \frac{(e_{2k+1}-e_{2k-1}) (10r_{2k-1}^2-5r_{2k-1}r_{2k}+r_{2k}^2) r_{2k}^3}{(r_{2k}-r_{2k-1})^5}\,,\\
\end{aligned}   
\\ \nonumber k\in\mathbb{N}\,, \quad k\in [1,K]\,. \hspace{-2.1cm} 
\end{align}

%-------------------------------------------------------------------------
\section*{Acknowledgements}
%-------------------------------------------------------------------------

We are grateful to C. Chirenti, F. Di Filippo, C. Ecker, S. Liberati,
K. Miler, and E. Mottola for insightful discussions and
comments. Partial funding comes from the State of Hesse within the
Research Cluster ELEMENTS (Project ID 500/10.006), by the ERC Advanced
Grant ``JETSET: Launching, propagation and emission of relativistic
jets from binary mergers and across mass scales'' (Grant No. 884631)
and by the Deutsche Forschungsgemeinschaft (DFG, German Research
Foundation) through the CRC-TR 211 ``Strong-interaction matter under
extreme conditions''-- project number 315477589 -- TRR 211. LR
acknowledges the Walter Greiner Gesellschaft zur F\"orderung der
physikalischen Grundlagenforschung e.V. through the Carl W. Fueck
Laureatus Chair.

%-------------------------------------------------------------------------

%-------------------------------------------------------------------------
\section*{References}

\bibliographystyle{iopart-num}

\begin{thebibliography}{10}
\expandafter\ifx\csname url\endcsname\relax
  \def\url#1{{\tt #1}}\fi
\expandafter\ifx\csname urlprefix\endcsname\relax\def\urlprefix{URL }\fi
\providecommand{\eprint}[2][]{\url{#2}}

\bibitem{Mazur2001}
{Mazur} P~O and {Mottola} E 2001 {\em arXiv e-prints\/} gr-qc/0109035
  (\textit{Preprint} \eprint{gr-qc/0109035})

\bibitem{Abbott2016fw}
Abbott B~P {\em et~al.\/} (Virgo, LIGO Scientific) 2016 {\em Phys. Rev.
  Lett.\/} {\bf 116} 241102 (\textit{Preprint} \eprint{1602.03840})

\bibitem{Akiyama2019_L1_etal}
{Akiyama} K and \textit{et al} 2019 {\em Astrophys. J. Lett.\/} {\bf 875} L1

\bibitem{EHT_SgrA_PaperI_etal}
{Akiyama} K and \textit{et al} 2022 {\em Astrophys. J. Lett.\/} {\bf 930} L12

\bibitem{Cardoso2019}
{Cardoso} V and {Pani} P 2019 {\em Living Reviews in Relativity\/} {\bf 22} 4
  (\textit{Preprint} \eprint{1904.05363})

\bibitem{Mazur2004}
{Mazur} P~O and {Mottola} E 2004 {\em Proceedings of the National Academy of
  Science\/} {\bf 101} 9545--9550 (\textit{Preprint} \eprint{gr-qc/0407075})

\bibitem{Schwarzschild16a}
Schwarzschild K 1916 {\em Sitzungsber. Dtsch. Akad. Wiss. Berlin, Kl. Math.
  Phys. Tech.\/} {\bf 1} 189--196

\bibitem{Rezzolla_book:2013}
{Rezzolla} L and {Zanotti} O 2013 {\em Relativistic Hydrodynamics\/} (Oxford,
  UK: Oxford University Press) ISBN 9780198528906

\bibitem{Sakharov1966}
{Sakharov} A~D 1966 {\em Soviet Journal of Experimental and Theoretical
  Physics\/} {\bf 22} 241

\bibitem{Gliner1966}
{Gliner} E~B 1966 {\em Soviet Journal of Experimental and Theoretical
  Physics\/} {\bf 22} 378

\bibitem{deSitter:1917zz}
de~Sitter W 1917 {\em Mon. Not. Roy. Astron. Soc.\/} {\bf 78} 3--28

\bibitem{Chirenti2007}
{Chirenti} C~B~M~H and {Rezzolla} L 2007 {\em Class. Quantum Grav.\/} {\bf 24}
  4191--4206 (\textit{Preprint} \eprint{0706.1513})

\bibitem{Mottola2023}
Mottola E 2023 {\em Gravitational Vacuum Condensate Stars\/} (Singapore:
  Springer Nature Singapore) pp 283--352 ISBN 978-981-99-1596-5
  \urlprefix\url{https://doi.org/10.1007/978-981-99-1596-5_8}

\bibitem{Dymnikova1992}
{Dymnikova} I 1992 {\em General Relativity and Gravitation\/} {\bf 24} 235--242

\bibitem{Dymnikova2000}
Dymnikova I 2000 Variable cosmological term - geometry and physics
  (\textit{Preprint} \eprint{gr-qc/0010016})

\bibitem{Dymnikova2005}
{Dymnikova} I and {Galaktionov} E 2005 {\em Class. Quantum Grav.\/} {\bf 22}
  2331--2357 (\textit{Preprint} \eprint{arXiv:gr-qc/0409049})

\bibitem{Mazur_2015}
Mazur P~O and Mottola E 2015 {\em Classical and Quantum Gravity\/} {\bf 32}
  215024 \urlprefix\url{https://doi.org/10.1088%2F0264-9381%2F32%2F21%2F215024}

\bibitem{Buchdahl:59}
Buchdahl H~A 1959 {\em Phys. Rev.\/} {\bf 116} 1027--1034

\bibitem{Carballo2018}
{Carballo-Rubio} R, {Di Filippo} F, {Liberati} S and {Visser} M 2018 {\em Phys.
  Rev. D\/} {\bf 98} 124009 (\textit{Preprint} \eprint{1809.08238})

\bibitem{Chirenti2016}
{Chirenti} C and {Rezzolla} L 2016 {\em Phys. Rev. D\/} {\bf 94} 084016
  (\textit{Preprint} \eprint{1602.08759})

\bibitem{Ray2020}
{Ray} S, {Sengupta} R and {Nimesh} H 2020 {\em International Journal of Modern
  Physics D\/} {\bf 29} 2030004-260

\bibitem{Chapline_2003}
Chapline G 2003 {\em International Journal of Modern Physics A\/} {\bf 18}
  3587--3590 \urlprefix\url{https://doi.org/10.1142%2Fs0217751x03016380}

\bibitem{Broderick2007}
{Broderick} A~E and {Narayan} R 2007 {\em Class. Quantum Grav.\/} {\bf 24}
  659--666 (\textit{Preprint} \eprint{arXiv:gr-qc/0701154})

\bibitem{EHT_SgrA_PaperVI_etal}
{Akiyama} K and \textit{et al} 2022 {\em Astrophys. J. Lett.\/} {\bf 930} L17

\bibitem{Cattoen2005}
{Cattoen} C, {Faber} T and {Visser} M 2005 {\em Class. Quantum Grav.\/} {\bf
  22} 4189--4202 (\textit{Preprint} \eprint{arXiv:gr-qc/0505137})

\bibitem{Mbonye2005}
{Mbonye} M~R and {Kazanas} D 2005 {\em Phys. Rev. D\/} {\bf 72} 024016
  (\textit{Preprint} \eprint{arXiv:gr-qc/0506111})

\bibitem{Carballo2022}
{Carballo-Rubio} R, {Di Filippo} F, {Liberati} S and {Visser} M 2022 {\em
  Journal of Cosmology and Astroparticle Physics\/} {\bf 2022} 055
  (\textit{Preprint} \eprint{2205.13555})

\bibitem{Carballo2023}
{Carballo-Rubio} R, {Di Filippo} F, {Liberati} S and {Visser} M 2023 {\em
  Journal of High Energy Physics\/} {\bf 2023} 46 (\textit{Preprint}
  \eprint{2211.05817})

\bibitem{Chirenti2008}
{Chirenti} C~B~M~H and {Rezzolla} L 2008 {\em Phys. Rev. D\/} {\bf 78} 084011
  (\textit{Preprint} \eprint{0808.4080})

\end{thebibliography}

\providecommand{\newblock}{}

%-------------------------------------------------------------------------

%==================================================================
%==================================================================

\end{document}